\documentclass[10pt, conference]{IEEEtran}

\usepackage{graphicx}

\usepackage[numbers,sort&compress]{natbib}

\usepackage{xurl}

\usepackage{tabularx}
\usepackage{booktabs}

\usepackage[inline]{enumitem}

\usepackage{braket}

\usepackage{tikz}
\usetikzlibrary{positioning,arrows.meta}

\newif\ifarxiv
\arxivtrue   %

\ifarxiv
    \usepackage[colorlinks]{hyperref}
    \usepackage{xcolor} %
    \def\tmp#1#2#3{%
      \definecolor{Hy#1color}{#2}{#3}%
      \hypersetup{#1color=Hy#1color}}
    \tmp{link}{HTML}{800006}
    \tmp{cite}{HTML}{2E7E2A}
    \tmp{file}{HTML}{131877}
    \tmp{url} {HTML}{8A0087}
    \tmp{menu}{HTML}{727500}
    \tmp{run} {HTML}{137776}
    \def\tmp#1#2{%
      \colorlet{Hy#1bordercolor}{Hy#1color#2}%
      \hypersetup{#1bordercolor=Hy#1bordercolor}}
    \tmp{link}{!60!white}
    \tmp{cite}{!60!white}
    \tmp{file}{!60!white}
    \tmp{url} {!60!white}
    \tmp{menu}{!60!white}
    \tmp{run} {!60!white}
\else
    \usepackage[hidelinks]{hyperref}
\fi

\newcommand{\apprefauto}[2]{%
  \ifarxiv
    Appendix~\ref{#1}%
  \else
    \cite[App.~#2]{miranskyy2026course}%
  \fi
}

\newcommand{\smcite}[2][]{%
  \ifarxiv
    \cite[#1]{#2}%
  \else
  \fi
}

\newcommand{\smcitet}[1]{%
  \ifarxiv
    \citet{#1}%
  \else
  \fi
}

\newcommand{\smlocation}[1]{%
  \ifarxiv
    Appendices provide detailed lecture preparation materials and representative in-class activities to support reproducibility.%
    Where particularly useful for course design, the paper cites \textit{specific sections or pages}; otherwise, references are intended to be read more broadly.
  \else
    Given space constraints, some implementation and instructional details are deferred to \emph{supplementary material} in~\cite{miranskyy2026course}. Where particularly useful for course design, the paper cites \textit{specific sections or pages}; otherwise, references are intended to be read more broadly.
  \fi
}

\title{A Course on the Introduction to Quantum Software Engineering: Experience Report}

\author{
\IEEEauthorblockN{Andriy Miranskyy}
\IEEEauthorblockA{\textit{Department of Computer Science, Toronto Metropolitan University} \\
Toronto, Canada \\
avm@torontomu.ca
}}

\begin{document}
\maketitle
\bstctlcite{BSTcontrol}

\begin{abstract}
Quantum computing is increasingly practised through programming, yet most educational offerings emphasize algorithmic or framework-level use rather than software engineering concerns such as testing, abstraction, tooling, and lifecycle management.

This paper reports on the design and first offering of a cross-listed undergraduate--graduate course that frames quantum computing through a software engineering lens, focusing on early-stage competence relevant to software engineering practice. The course integrates foundational quantum concepts with software engineering perspectives, emphasizing executable artifacts, empirical reasoning, and trade-offs arising from probabilistic behaviour, noise, and evolving toolchains. Evidence is drawn from instructor observations, supplemented by anonymous student feedback, a background survey, and inspection of student work.

Instructor observations and inspected work artifacts indicated that students with little prior exposure could engage with quantum software engineering topics after foundational quantum concepts had been introduced using executable artifacts. The contribution is the systematic software engineering framing and sequencing of the course, its mixed-level assessment design, and practice-grounded lessons from an actual offering; the report does not estimate causal learning effects.

\end{abstract}

\begin{IEEEkeywords}
quantum software engineering,
curriculum development,
course design,
experience report
\end{IEEEkeywords}

\maketitle

\section{Introduction}
\label{sec:intro}

Quantum computing has progressed from a largely theoretical research area to an emerging, software-intensive discipline~\cite{murillo2025quantum}. Cloud-accessible quantum platforms and open-source programming frameworks now allow developers to write, execute, and experiment with quantum programs in ways that closely resemble classical software development~\cite[Sec. 4.2]{murillo2025quantum}. As a result, quantum programs are increasingly treated as software artifacts rather than purely mathematical abstractions, amplifying the need for developers who can reason about quantum computing through a software engineering (SE) lens~\cite[pp. 17--18]{murillo2025quantum}.

The course was motivated by three converging signals that have emerged from distinct aspects of the author's work since 2018, namely teaching, research supervision, and industry collaboration.
In instructional settings, students were often able to execute quantum programs using existing libraries but struggled to reason about correctness, testing, and evolution once they moved beyond small, illustrative circuits. Similar patterns appeared in research supervision: graduate students with strong SE backgrounds found it difficult to apply familiar testing, abstraction, or design principles without a clear operational understanding of what quantum code was doing. Industry collaborators reported analogous challenges, noting that while quantum functionality could often be accessed through well-defined APIs, the internal structure of quantum programs remained opaque, making inspection, modification, and maintenance difficult even for experienced software engineers.

These challenges are becoming more salient as vendor roadmaps describe quantum hardware and software systems of increasing scale and complexity. For example, IBM and Quantinuum each target fault-tolerant systems by 2029~\cite{mandelbaum2025scaling,mandelbaum2025ibm,quantinuum2025quantinuum}. This trajectory strengthens the need to treat quantum programs as software systems subject to testing, abstraction boundaries, tooling trade-offs, and lifecycle concerns~\cite{Piattini21-IT-pro,ali2022software,zhao2020quantum}.

Despite rapid growth in quantum computing education, most existing university-level courses remain algorithm- or theory-centric. Programming is commonly introduced in a style resembling scientific computing, where students construct short scripts to build circuits and observe outcomes, but rarely confront issues of testing, design structure, or long-term maintainability~\cite{watrous2018theory,watrous2025understanding,wong2022introduction}. Based on the author's experience, incremental curricular adjustments have proven insufficient. Adding a single lecture on quantum computing to an SE course does not provide adequate exposure for meaningful competence, while incorporating a short SE module into a quantum computing course would require substantial instructional time that is difficult to accommodate. These constraints motivate the need for a dedicated course that systematically frames quantum computing as an SE problem. Unlike algorithm-centric quantum computing courses~\cite{watrous2018theory,watrous2025understanding,wong2022introduction} or framework-centric programming tutorials~\cite{qiskitextbook2023,pennylane_codebook}, the course described in this paper treats quantum programs as evolving software systems, emphasizing core software system attributes under quantum-specific constraints.
In this paper, framing quantum computing as an SE problem means treating quantum programs as executable software artifacts subject to concerns such as abstraction, testing, tooling, platform variability, maintainability, migration, and lifecycle reasoning, rather than presenting them solely as mathematical objects or algorithmic results.

This paper reports on the design and delivery of a cross-listed undergraduate--graduate elective that introduces quantum computing explicitly as an SE domain. The course targets onboarding-level competence rather than mastery. Students are not expected to become experts in quantum algorithms; instead, they are expected to develop the ability to reason about quantum programs as software artifacts, including their abstractions, tooling choices, empirical behaviour, and evolution. As an introductory offering, the course emphasizes SE concerns that are especially visible in early quantum software practice, such as abstraction, testing, tooling, and lifecycle issues. Topics such as requirements engineering and software architecture are therefore positioned as important directions for future expansion. This course is not intended to replace mathematically rigorous quantum computing courses; rather, it aims to provide enough mathematical\footnote{
In particular, the foundational quantum-information component relied on mathematically substantive course materials by \citet{watrous2025understanding} rather than intuition alone, as discussed in \apprefauto{sec:lecture_details}{A}.
} and conceptual foundation for software engineering students to reason meaningfully about quantum programs within the bounded scope of a single elective.

The course was offered as a mixed-level elective in the Winter 2025 term.
The evidence presented in this paper is observational and experience-based, drawing on instructor reflections, student feedback\footnote{Representative formative feedback instruments are shown in \apprefauto{sec:feedback_forms}{E}.}, and analysis of student work.
The goal is to support adoption and adaptation of similar courses rather than to provide controlled measurements of learning outcomes.
The course was designed and taught by the author, and all instructor observations reported in this paper derive from direct, first-hand involvement in course design, delivery, and assessment.

The novelty claimed here is not any individual quantum topic, framework, testing method, or use of project-based learning. It is their systematic SE framing, the sequence from executable quantum foundations into QSE, the mixed-level assessment design, and the lessons documented from an actual offering. The paper \textit{contributes}:
\begin{enumerate*}
  \item a documented course design that integrates quantum computing fundamentals with SE concerns such as abstraction, testing, tooling, and lifecycle issues,
  \item a modular pedagogical structure that leverages research roadmaps to organize quantum software engineering (QSE) topics and connect them to executable artifacts and classroom activities,
  \item an assessment model suitable for mixed undergraduate and graduate audiences, and
  \item context-sensitive observations and lessons for instructors considering similar offerings.
\end{enumerate*}

The remainder of the paper is organized as follows. Section~\ref{sec:background} situates the course within the broader landscape of quantum computing education and identifies gaps related to SE concerns. Section~\ref{sec:course} describes the course design and delivery, including its objectives, lecture structure, and assessment components. Section~\ref{sec:students} characterizes the student population and course dynamics, while Section~\ref{sec:lessons} synthesizes observations and lessons from the first offering. Section~\ref{sec:threats} discusses threats to validity and limitations, and Section~\ref{sec:summary} concludes with directions for future refinement and replication.
\smlocation{}

\section{Background and Educational Context}
\label{sec:background}

This section situates the course within the broader landscape of quantum computing education. It outlines common educational pathways, highlights gaps in software engineering (SE), and motivates the need for a dedicated QSE perspective.

Most university-level quantum computing courses prioritize quantum information theory and algorithms. Canonical offerings, such as those by~\citet{watrous2018theory,watrous2025understanding}, provide conceptual clarity and mathematical rigour, making them well suited for students seeking principled introductions to quantum computation. Textbooks such as~\citet{wong2022introduction} adopt a computer-science-oriented perspective and are accessible to advanced undergraduates, while more comprehensive references such as~\citet{nielsen_chuang_2010} are regarded as mathematically demanding for most SE undergraduate audiences. Across these resources, the emphasis remains on foundational concepts and algorithmic models rather than on software lifecycle concerns.

The closest related work is \citet{haghparast2024innovative}, which presents a multi-framework agenda for teaching quantum programming and SE, covering quantum foundations, software-development lifecycle topics, containerization, and practical use of Qiskit, PennyLane, and Ocean. 
That work is complementary rather than overlapping: rather than proposing a tooling-focused agenda, this paper reports on the design and first offering of a full course that devotes more time to establishing the quantum-computing background needed for students to reason effectively about subsequent SE topics.

Other recent works are likewise complementary. \citet{carberry2024peer} focus on a peer-instruction board game for introducing quantum technologies to engineering students; \citet{lukishova2024teach} provides a broad overview of quantum education across labs, curriculum development, outreach, and large-enrollment teaching; and \citet{artner2023introducing} present a rogue-like educational game helping computer science (CS) students understand the mathematics and execution of quantum computations. In contrast, this paper offers a course-level experience report on structuring, delivering, and reflecting on a SE-oriented introductory course for mixed  audiences.

In parallel, vendor- and community-driven resources (such as platform-specific training portals like the PennyLane Codebook~\cite{pennylane_codebook} or the now deprecated Qiskit Textbook~\cite{qiskitextbook2023}) support hands-on experimentation and tool familiarity. While valuable for learning specific frameworks, these materials are typically self-guided, platform-centric, and not designed as comprehensive academic curricula. As a result, they offer limited support for reasoning about design trade-offs, testing strategies, or long-term maintainability of quantum software.

Prior work highlights structural limitations in quantum computing education. A 2020 empirical study of practitioners~\cite{shaydulin2020making} reported few formal training pathways, with most contributors relying on self-directed learning and ad hoc combinations of physics, mathematics, and CS backgrounds. In 2025, a comprehensive roadmap of QSE research~\cite[pp.~17--18]{murillo2025quantum} identified workforce training as a prerequisite and a challenge for the field's maturation, emphasizing the need to define comprehensive training strategies and develop systematic QSE methodologies to support the transition from classical to quantum software development. Together, these findings suggest that existing educational offerings lag behind emerging QSE practice and have yet to systematically adopt an SE framing.

The absence of an explicit SE perspective leaves several gaps. Students are rarely exposed to testing, debugging, and quality assurance practices for quantum programs, despite nondeterminism, noise, and probabilistic outcomes being intrinsic to current hardware. Lifecycle concerns (such as requirements engineering, abstraction design, tooling choices, backend portability, and maintainability) are often implicit or omitted entirely.

From an SE perspective, quantum computing allows limited reuse of classical techniques but also breaks key assumptions on which many SE methods rely~\cite{miranskyy2019testing,murillo2025quantum}. Determinism, repeatability, and clear test oracles are foundational in classical SE, yet quantum programs exhibit inherent probabilistic behaviour and sensitivity to noise and platform variability~\cite{miranskyy2019testing,murillo2025quantum}. For example, quantum programs cannot be debugged in the classical sense on real quantum hardware: the internal quantum state is not directly observable, and repeated executions yield only statistical distributions~\cite{miranskyy2020quantum,miranskyy2021testing}. These mismatches complicate direct transfer of established SE techniques and necessitate careful adaptation or the development of new approaches. Addressing these tensions requires educational experiences that explicitly prepare students to reason about quantum programs as evolving software systems rather than as isolated algorithmic artifacts.

This gap motivates the design of a dedicated course that frames quantum computing explicitly through an SE lens, bridging foundational quantum concepts with practical concerns across the software lifecycle.

\section{Course Design and Delivery}
\label{sec:course}

This section describes the design and delivery of the course, focusing on its educational goals, instructional structure, and assessment model. 
It first outlines the target competencies that guided the course design (Section~\ref{subsec:goals}). It then describes the lecture structure and pedagogical sequencing (Section~\ref{subsec:lecture}), followed by the assessment components: assignments (Section~\ref{sec:assignments}), examinations for undergraduate students (Section~\ref{sec:exam}) and research paper presentations for graduate students (Section~\ref{sec:research_paper}), the course project (Section~\ref{sec:project}), and in-class participation (Section~\ref{sec:participation}). Table~\ref{tab:assessment-weighting-breakdown} summarizes the relative weighting of these components.

\begin{table}[tb]
  \centering
  \caption{Assessment weighting by component.}
  \label{tab:assessment-weighting-breakdown}
  \begin{tabular}{@{}p{6.0cm} >{\raggedleft\arraybackslash}p{1.5cm}@{}}
    \toprule
    Component & Weight \\
    \midrule
    Assignments (Section~\ref{sec:assignments}) & 30\% \\
    Final exam for undergrad (Section~\ref{sec:exam}) or research paper presentation for grads (Section~\ref{sec:research_paper}) & 20\% \\
    Project (Section~\ref{sec:project}) & 45\% \\
    In-class participation (Section~\ref{sec:participation}) & 5\% \\
    \bottomrule
  \end{tabular}
\end{table}

\subsection{Educational Objectives and Target Competencies}
\label{subsec:goals}

The course was designed to enable students to reason about quantum computing as an SE problem domain. Rather than treating quantum programs solely as mathematical or algorithmic artifacts, the course framed them as executable software systems subject to nondeterminism, noise, tooling constraints, and lifecycle considerations. The course was positioned as an advanced SE elective within a standard CS curriculum and was intended to complement, rather than replace, foundational quantum computing courses. Foundational material in quantum information and quantum algorithms was taught through executable and inspectable representations specifically to enable later reasoning about software lifecycle concerns and related software system attributes introduced in Section~\ref{sec:intro}.

More concretely, the course framed quantum computing as an SE problem in four recurring ways. First, students were asked to interpret quantum programs as executable artifacts rather than only as formal derivations. Second, lecture and practicum activities emphasized abstraction boundaries, inspectability, and black-box versus white-box reasoning. Third, students engaged with SE challenges that arise from noise, nondeterminism, backend variability, and immature tooling, especially in relation to testing, debugging, and empirical evaluation. Fourth, topics such as post-quantum cryptography (PQC) and programming paradigms were used to connect quantum computing to broader concerns of migration, maintainability, language design, and lifecycle evolution.

At the undergraduate level, the \textit{primary objective} was onboarding-level competence. By the end of the course, students were expected to 
\begin{enumerate*}
    \item write and execute basic quantum programs,
    \item understand how quantum states, circuits, and measurements are represented in code,
    \item reason about probabilistic execution and empirical results, and
    \item articulate SE challenges arising from current quantum platforms, including testing, debugging, abstraction design, and backend variability.
\end{enumerate*}
Emphasis was placed on breadth of exposure to QSE challenges rather than depth in any single subdomain.

A \textit{secondary objective} was the development of tool fluency. Tool fluency was understood not only as the ability to use specific frameworks, but also as the ability to reason critically about tooling choices, abstractions, and trade-offs across simulators, programming frameworks, and hardware backends. Students were expected to compare execution environments, interpret noisy or inconclusive results, and avoid overclaiming correctness based on limited empirical evidence.

For graduate students, the course included an \textit{additional objective} of developing early research skills. Graduate students were expected to read and analyze contemporary research literature in QSE, identify limitations or open challenges, and reason about possible extensions or alternative approaches. These competencies were reinforced through research paper presentations and open-ended project work.

Foundational quantum computing material was introduced only to the extent necessary to support these goals. Deep mathematical derivations, exhaustive algorithmic coverage, and hardware-level detail were intentionally de-emphasized in favour of core software system attributes and empirical reasoning.

The course was designed for a mixed undergraduate and graduate audience in CS. For undergraduate students, \textit{formal prerequisites} included linear algebra, probability and statistics, computer organization, introduction to SE, algorithms, and computer security. These prerequisites reflect the interdisciplinary nature of QSE, which draws on foundations from mathematics, systems, and SE.

No prior experience with quantum computing was expected. To accommodate heterogeneous backgrounds, foundational quantum concepts were introduced incrementally, with an emphasis on executable artifacts and code-level representations rather than mathematical formalism. Prior exposure to advanced SE topics was beneficial but not required.

\subsection{Lectures}
\label{subsec:lecture}

The course comprised twelve three-hour lectures for a mixed undergraduate and graduate audience. The extended blocks were intended to sustain focus on a single topic while accommodating theory, discussion, and hands-on work within one meeting; each used three 50-minute segments separated by 10-minute breaks. Practica and graduate presentations typically occupied the latter part of a meeting. During this first offering, practica remained in the lecture so the instructor could observe difficulties and refine activities before possible migration to teaching assistant (TA)-led labs. Preparatory materials and activities are detailed in \apprefauto{sec:lecture_details}{A} and \apprefauto{sec:participation_activities}{B}.

Lecture topics were organized to first establish a conceptual foundation in quantum computing before introducing SE concerns.  This sequencing was intentional: early emphasis on executable, inspectable representations was used to reduce abstraction barriers and enable later reasoning about software lifecycle issues once students had developed an operational understanding of quantum program behaviour. The course began with foundational quantum information concepts, followed by canonical quantum algorithms, and then transitioned to QSE topics (Table~\ref{tab:lecture_se_mapping}). This sequencing was intended to mitigate the abstraction gap commonly encountered by CS students when first exposed to quantum computing.

Quantum algorithms were introduced as computational systems whose structure and execution could be inspected and reasoned about, rather than as opaque mathematical constructs. This ``white-box'' framing emphasized executable artifacts and operational understanding. Once students had developed sufficient phenomenological understanding to reason about quantum programs, the course shifted to QSE topics spanning quality assurance, service-oriented computing, and programming paradigms.

In-class observations and student comments consistently identified the early quantum information and algorithms material as more difficult than the later QSE topics, which connected more directly to prior SE coursework. These observations guided pacing but were not a comparative learning-outcomes measure.

A flipped classroom approach~\cite{maher2015flipped,green2020flipped} was employed selectively. For Lectures~2--4 and~7--9, students used pre-recorded videos and readings before class and completed short, automatically graded quizzes at the start of the lecture. Collective review of the results exposed misconceptions and informed pacing and discussion. Other lectures used a traditional format when suitable preparation was unavailable. Anonymous responses were used for formative calibration, not to estimate the effect of flipping; the instruments are reproduced in \apprefauto{sec:feedback_forms}{E}.

Foundational quantum concepts drew on openly available instructional materials, supplemented by in-class explanations that mapped abstract operations (such as quantum gates) to their representation in code (\apprefauto{sec:lecture_details}{A}). Quantum algorithms were presented at varying levels of abstraction, reflecting common SE practice in which developers reason about system behaviour without complete theoretical mastery.

In the QSE portion of the course, lectures were structured around the research themes presented in the QSE roadmap~\cite{murillo2025quantum}. This survey provided an organizing structure for introducing QSE subdomains and for selecting representative research papers as entry points for deeper discussion. Rather than attempting exhaustive coverage, the course emphasized understanding classes of problems and open challenges across the software lifecycle.

Beginning in Lecture~2, selected lectures incorporated short in-class practica that translated abstract ideas into executable artifacts and exposed empirical variability, noise, and tooling constraints. A representative implementation sequence in Lecture~3 began with students implementing quantum teleportation from a reference in a circuit-based framework. They then exported the circuit to OpenQASM and imported it into a second framework. This workflow exposed architectural and language-level differences, including limitations in real-time classical control in some toolchains used at the time of the course. Students finally implemented teleportation natively in the second framework and compared language expressivity, abstraction choices, and representational constraints. The runnable algorithm was therefore a vehicle for reasoning about interoperability and portability rather than the endpoint of the activity.

Lecture~4 extended this emphasis from representation to execution. Students measured Bell states on an ideal simulator, a noisy simulator, and real quantum hardware, then compared the observed distributions across backends. They also inspected the source of a packaged Shor implementation from the deprecated Qiskit Aqua library, connecting empirical execution with library evolution and API lifecycle concerns. In Lecture~8, groups inspected a defective quantum program, identified defects, proposed tests and fixes, and compared debugging strategies during the debrief. That activity made test-oracle and observability problems concrete. Complete descriptions of these and the other practica appear in \apprefauto{sec:participation_activities}{B}.

All hands-on activities were conducted using browser-based Jupyter notebooks~\cite{loizides2016jupyter} (via Google Colab~\cite{google_colab_online}), reducing technical barriers and eliminating concerns about local environment configuration. Students implemented and executed quantum programs using widely adopted open-source frameworks, including OpenQASM~\cite{cross2022openqasm}, PennyLane~\cite{bergholm2018pennylane}, and Qiskit~\cite{javadi2024quantum}, exposing them to both circuit-based and hybrid quantum-classical programming models. As part of their course projects, students were also encouraged to explore additional frameworks of their choice, allowing comparison of alternative abstractions and ecosystems.

\begin{table*}[t]
\centering
\footnotesize
\setlength{\tabcolsep}{3pt}
\caption{Mapping of lecture topics to software engineering concerns. The lecture-by-lecture preparation guide appears in \apprefauto{sec:lecture_details}{A}.}
\label{tab:lecture_se_mapping}
\begin{tabular}{@{}p{1.3cm}p{3.5cm}p{4.8cm}p{6.0cm}@{}}
\toprule
Lecture(s) & Topic & Main SE concerns & Illustrative focus \\
\midrule
1 & Introduction and motivation & Software lifecycle impact, maintainability, legacy-system risk, emerging engineering challenges & Quantum computing framed as a software-intensive domain affecting both new quantum software and classical systems exposed to quantum-era risks \\
\addlinespace
2--3 & Quantum information and circuits & Executable representations, abstraction, inspectability, reasoning about system state and behaviour & Qubits, gates, circuits, and teleportation introduced through code-level and circuit-level representations rather than only formal mathematics \\
\addlinespace
4--5 & Quantum algorithms & Abstraction boundaries, white-box vs.\ black-box reasoning, implementation structure, reuse, engineering trade-offs & Grover's and Shor's algorithms discussed as executable software artifacts at carefully chosen levels of abstraction \\
\addlinespace
6 & Post-quantum cryptography & Migration, backward compatibility, standards adoption, risk management, long-term system evolution & PQC presented as an SE problem of system transition under changing standards and threat models \\
\addlinespace
7 & Scope and foundations of QSE & Software development lifecycle, problem decomposition, research roadmap, field structuring & QSE introduced as a set of challenges and practices spanning the lifecycle rather than as an isolated technical topic \\
\addlinespace
8 & Quality assurance and service-oriented computing & Testing, debugging, oracle problems, platform variability, service abstraction & Quantum software quality assurance and service-oriented concerns explored through representative challenges and research directions \\
\addlinespace
9 & Programming paradigms & Language design, abstraction mechanisms, expressivity, tool support & Comparison of programming paradigms with emphasis on quantum-native abstractions and higher-level language support \\
\addlinespace
10 & Future of quantum computing & Technology forecasting, planning under uncertainty, long-term engineering adaptation & Historical and forward-looking discussion used to connect uncertainty in the field to planning and design choices in software systems \\
\addlinespace
11 & Recap and discussion & Integration and synthesis & Consolidation of links between quantum foundations, algorithms, and software-engineering concerns \\
\addlinespace
12 & Project presentations & Open-ended design, empirical evaluation, communication of engineering results & Student projects used to synthesize implementation, experimentation, and QSE-oriented reflection \\
\bottomrule
\end{tabular}
\end{table*}

\subsection{Assignments}
\label{sec:assignments}

The course included two individual assignments, together accounting for 30\% of the final grade, with intentionally conservative weighting. At the time of course delivery, the rapid adoption of large language models (LLMs) in programming and technical writing raised unresolved questions about how to design robust, fair, and meaningful assessments in CS and SE education~\cite{daun2023chatgpt,kirova2024software}. Rather than attempting ad hoc mitigation strategies, which remain an open research and pedagogical problem~\cite{raihan2025large}, the course adopted a transparent assessment design that emphasized conceptual synthesis and critical reasoning, making individual work meaningful regardless of tool assistance. Students were explicitly informed that LLMs could be used, but only critically, and that individual assignments were weighted modestly given the evolving state of best practices for assessment in the presence of such tools.

The assignments served two complementary purposes: 
\begin{enumerate*}
    \item to ensure a baseline level of conceptual readiness in quantum computing fundamentals and
    \item to encourage conceptual synthesis and critical reasoning about QSE topics introduced in lectures.
\end{enumerate*}
Representative prompts for both assignments are provided in \apprefauto{sec:assessment_materials}{C}.

The first assignment was designed as a skills-bootstrapping exercise. It consisted of two components. The first required students to pass an externally administered fundamentals test on quantum information offered as part of~\cite{watrous2025understanding}. The test comprised 20 multiple-choice questions, with a minimum threshold of 16 correct answers required to pass. This component ensured a common baseline of conceptual understanding while minimizing duplication of instructional effort within the course.

The second component focused on classical data encoding techniques for quantum computation, specifically basis encoding and amplitude encoding~\cite{minati2024quantum}. Students implemented small-scale examples, constructed corresponding quantum circuits, and answered conceptual questions probing the scalability and limitations of these approaches. Although the assignment did not require implementing full quantum machine learning algorithms, the encoding tasks were intentionally selected to expose a practical bottleneck in quantum computing: loading classical data into quantum states~\cite{miranskyy2024comparing}.

One representative prompt supplied the bitstrings \texttt{0101} and \texttt{1011} and required code plus images of both the state-preparation and transpiled circuits. A follow-up scenario asked how ten 128-bit strings could be handled on a computer with 16GB of RAM, directing attention to sparse representations rather than an infeasible dense statevector. A second prompt supplied the values $[2,7,6,2,5]$ for amplitude encoding, required state-preparation and transpiled circuits using \textsc{rx}, \textsc{ry}, \textsc{rz}, and \textsc{cx} as basis gates, and asked how negative values could be represented without losing sign information. These prompts coupled implementation artifacts with resource and representation reasoning; the complete wording appears in \apprefauto{sec:assessment_materials}{C}.

Interest in quantum machine learning was identified early in the term through an anonymous survey administered during the first lecture, and the design of this component partially responded to that interest. Several students informally reported that the encoding exercises helped them appreciate challenges related to memory requirements, sparsity, and circuit growth when working with realistic datasets. For students who later pursued course projects related to quantum machine learning, this assignment provided an accessible entry point without requiring early commitment to full algorithmic pipelines.

The second assignment was an analytical QSE synthesis. Students submitted short essays addressing topics such as PQC transition timelines, evolving definitions and scope of QSE, and trade-offs between idealized simulators and execution on real quantum hardware. For example, one prompt required a comparison of simulators and real hardware as software-testing environments, including the effects of noise, resource constraints, and scalability. The questions were explicitly aligned with the PQC and QSE lectures and required students to synthesize material across multiple sessions rather than reproduce isolated facts or code fragments. Taken together, these prompts required students to analyze quantum computing topics through SE concerns such as migration, testing strategy, abstraction, and language design.

Grading for this assignment was holistic, emphasizing clarity of argumentation, technical correctness, and the ability to connect SE concerns with quantum-specific constraints. This format was selected both to promote deeper engagement with the literature and to reduce reliance on narrowly structured responses that may be easily automated.

All students successfully completed both assignments; no inferential comparison of undergraduate and graduate performance was conducted. %

\subsection{Final Exam (Undergraduate Students Only)}
\label{sec:exam}

Undergraduate students were assessed using a final exam, as required by university regulations. In addition to satisfying this institutional constraint, the exam provided a clear mechanism for differentiating assessment expectations between undergraduate and graduate students, whose learning objectives and evaluation methods differed.

The final exam accounted for 20\% of the overall course grade and was designed to assess conceptual understanding of quantum information, the ability to reason about quantum programs, and the capacity to interpret quantum software artifacts through an SE lens. All material covered in lectures was considered in scope.

The exam consisted of approximately twenty multiple-choice questions. Representative examples are provided in \apprefauto{sec:exam_examples}{F}. Questions covered definitions, short code and circuit interpretation, measurement outcomes, and SE topics such as testing under noise, low-level abstractions, and PQC migration risks. Although answers were selected from predefined options, many questions required tracing circuit behaviour or reasoning about amplitudes. Multiple-choice questions were chosen to reduce grading ambiguity in a first offering.

The exam was two hours in duration and was conducted in person under proctored, closed-book conditions. No undergraduate student failed it. Graduate students instead completed research paper presentations, reflecting their distinct learning objectives (Section~\ref{sec:research_paper}).

\subsection{Research Paper Presentations (Graduate Students Only)}
\label{sec:research_paper}

Graduate students were assessed through individual research paper presentations rather than a final exam. This assessment modality reflected the distinct learning objectives for graduate students and emphasized engagement with current research, critical analysis, and synthesis of emerging ideas in quantum computing and QSE.

The graduate assessment consisted of two components: paper selection (2\% of the final grade) and the presentation itself (18\%). Students were required to propose a research paper for presentation, subject to instructor approval. Along with the citation, students submitted a brief written justification explaining the relevance and suitability of the selected paper. Proposals were evaluated using a structured rubric capturing relevance to the course, novelty, scientific rigour, feasibility, and potential for class engagement. This process ensured both student agency and alignment with course objectives.

Presentations were delivered individually during regular lecture sessions and were limited to 15 minutes, followed by approximately 10 minutes of moderated discussion. Undergraduate students were present during these sessions, and presenters were expected to make the material accessible to a mixed undergraduate--graduate audience while preserving technical depth. Where feasible, presentation topics were scheduled to align loosely with the lecture sequence, although exact alignment was not always possible given diverse student interests.

Students were instructed to address the motivation and importance of the topic, the paper's objectives, its methodology, and key results. Evaluation was guided by a detailed rubric assessing oral delivery, clarity of explanation, adequacy of background context, relevance to CS and SE, handling of questions, and quality of visual materials. Feedback was provided in written form and served a primarily summative role.

The presentation assessment differentiated graduate expectations from the undergraduate exam. It fit this cohort when distributed across several weeks; larger graduate cohorts would likely require group presentations, parallel tutorials, or asynchronous recordings with rubric-based feedback.

\subsection{Course Project}
\label{sec:project}
The course project served as the primary integrative assessment and accounted for 45\% of the final grade, distributed across three components: a project idea/proposal (4.5\%), an in-class project presentation (4.5\%), and the final written project report (36\%).
The project was shared by undergraduate and graduate students, with identical formal expectations. It was designed to support three complementary learning objectives:
\begin{enumerate*}
\item applying quantum programming concepts in practice, 
\item exploring QSE challenges, and 
\item enabling open-ended exploration and creativity.
\end{enumerate*}
Together, these objectives encouraged students to engage with quantum computing as a software-intensive discipline rather than as a collection of isolated techniques.
From an SE perspective, the project functioned as the main setting in which students had to confront realistic questions of tooling choice, implementation structure, reproducibility, evaluation, and scope management.

Projects were conducted primarily in groups of up to three students, although individual projects were permitted. Mixed undergraduate--graduate teams were allowed and encouraged. Early in the term, students participated in an informal idea-pitching and team-formation process. In practice, graduate students tended to gravitate toward more complex or research-oriented problem framings, while undergraduate students gravitated toward more exploratory and empirical project work.

The project lifecycle included two main deliverables: a project proposal and a final report. Proposals were due early in the term and took the form of a structured abstract, including background, objectives, methods, expected results, conclusions, and implications. This format encouraged early consideration of scope and feasibility and provided an opportunity for formative instructor feedback. Proposals were reviewed and, when necessary, iteratively refined before approval.

Projects were implementation-focused and empirical in nature. Students were free to choose topics and tools, subject to instructor approval, and no constraints were imposed on programming languages or platforms\footnote{Students were also encouraged to experiment with additional quantum programming languages and frameworks of their choice (e.g., Qmod~\cite{vax2025qmod} and Qrisp~\cite{seidel2024qrisp}), allowing them to explore alternative abstractions and ecosystems beyond those used in the core lectures.}. In practice, many projects relied on simulator-based execution using widely adopted frameworks. Representative project themes included comparisons of quantum and classical algorithms, applications of quantum computing to optimization and finance, quantum machine learning, feasibility studies of algorithms like QAOA~\cite{farhi2014quantum}, and tooling-oriented investigations examining reliability, debugging, or developer support for quantum software. Students were explicitly asked to reflect on SE challenges encountered during their projects, such as testing, performance evaluation, experimental reproducibility, and tooling limitations.

The final deliverable was a written project report (maximum ten pages, using the ACM Primary Article Template~\cite{acm_template}), excluding references and appendices. Reports were expected to describe methodology in sufficient detail to support reproducibility and to analyze results rather than merely report outcomes. Evaluation considered technical depth, organization, conclusions and future work, supporting material, originality, and presentation of visuals or artifacts. Projects with software artifacts also submitted source code and a short \texttt{README.md}; the complete deliverables and criteria appear in \apprefauto{sec:project_appendix}{D}. All teams delivered a short final-lecture presentation, sometimes with a live demonstration.
Instructor support included proposal vetting, periodic check-ins, and office hours.

In larger offerings, proposal review, milestone tracking, and rubric-guided assessment can be distributed across trained TAs, while synchronous final presentations can be replaced by asynchronous videos, poster sessions, or parallel lab demos.

The projects provided sustained implementation work, and one was subsequently extended into a peer-reviewed research paper. This is a descriptive example of follow-on work, not evidence of a general project effect.

\subsection{Participation and In-Class Activities}
\label{sec:participation}
Participation was evaluated through structured in-class activities conducted during Lectures~2 through~8. The primary goals of this component were to encourage attendance, sustain engagement during extended three-hour lectures, and support peer learning through discussion and collaboration. Participation was explicitly framed to students as a learning mechanism first and an assessment mechanism second.

Each selected lecture included a tightly scripted activity lasting approximately 30 to 50 minutes. Students formed ad hoc groups with peers seated nearby, resulting in varying group compositions across lectures. Activities reinforced recently introduced material through short coding tasks, conceptual reasoning or prediction exercises, and debugging-oriented ``bug hunts''. They emphasized active engagement and exploration rather than polished artifacts, consistent with active-learning approaches reported in SE education~\cite{garcia-holgado2018pilot,maxim2019student}.

At the conclusion of each activity, groups were required to submit a small deliverable through the course learning management system. Submissions typically consisted of short written responses or screenshots of results. Deliverables were graded on a credit/no-credit basis and evaluated solely for effort and completion rather than correctness. Participation marks were evenly distributed across the relevant lectures and collectively accounted for 5\% of the final course grade. Given the low overall weight and binary grading scheme, this component was intentionally non-punitive.

Each activity concluded with a structured group debrief. During these debriefs, groups shared their results and discussed challenges encountered during the activity. Class-wide discussion focused on comparing alternative solution paths, surfacing common misconceptions, and reflecting on trade-offs, rather than identifying a single correct answer. This format supported reflective discussion and peer learning while normalizing partial or evolving understanding.

For larger offerings, these activities could be tested as TA-led tutorials or labs, using brief deliverables and rubric-guided checking. This adaptation was not evaluated in the first offering and would depend on access to teaching assistants with sufficient quantum computing and QSE preparation. Representative activities appear in \apprefauto{sec:participation_activities}{B}.

\section{Student Population and Course Dynamics}
\label{sec:students}

This course was offered as a cross-listed undergraduate and graduate elective within the Department of Computer Science. The elective format likely selected for students already interested in the topic. Initial enrolment comprised 27 undergraduate, 9 MSc, and 4 PhD students; by the end of the term, 24 undergraduate, 7 MSc, and 3 PhD students remained, all enrolled in CS programs. Graduate withdrawals occurred early, before the transcript deadline, whereas the three undergraduate withdrawals occurred later. As mentioned in Section~\ref{sec:intro}, the mixed undergraduate--graduate cohort was intentional, with assessment structures designed to accommodate heterogeneous academic levels. These counts describe cohort flow only: the small graduate denominators and absence of documented withdrawal reasons preclude comparative retention claims or explanations for withdrawal.

Among the undergraduate completers, 18 of 24 ($75\%$) were in their fourth year; the few junior students were typically transfer students with relevant foundational coursework from other institutions. The course listed several formal prerequisites for undergraduate students (shown at the end of Section~\ref{subsec:goals}). A small number of prerequisite waivers were granted based on academic records or documented industrial experience, most commonly for algorithms and computer security, and all students granted waivers completed the course. An anonymous first-lecture survey asked one binary question about prior quantum computing experience; two of 24 respondents (one undergraduate and one graduate) reported such experience. Seventeen of the 24 undergraduate completers ($\approx 71\%$) had completed co-op placements or internships. These background indicators are descriptive: the course did not test whether year level, waivers, prior quantum exposure, or industry experience affected preparedness, behaviour, or learning, and the cohort composition limits inference about earlier-year undergraduates.

Instructor observations included student comfort with unfamiliar code and with testing and debugging, greater analytical depth in some graduate discussions and graduate-only projects, and stable participation across the term. Each practicum required a short in-class deliverable and included group discussion, creating regular opportunities for involvement and informal peer interaction. No common assessment or comparison condition supported attributing these observations to academic level, industry experience, or the activities themselves; they are therefore reported as course dynamics rather than measured differences or effects.

These background characteristics delimit the context for the observations reported below.

\section{Observations and Lessons Learned}
\label{sec:lessons}

This section synthesizes context-sensitive observations rather than general prescriptions. Evidence came from instructor observations during lectures and project supervision; anonymous feedback on pacing, clarity, and activities (\apprefauto{sec:feedback_forms}{E}); the entry survey; and inspection of assignments, exam performance, presentations, and project reports. These sources were used descriptively, not to support causal claims. Interpretation must account for the elective format, mixed academic levels, and limited prior quantum experience.

\subsection{What Worked Well}
\label{subsec:what-worked}

The progression from quantum fundamentals to algorithms and then QSE topics (Section~\ref{subsec:lecture}) was workable in this offering. Instructor observations and student comments identified the initial quantum material as an abstraction barrier and the later SE discussions as comparatively accessible after students had used states and circuits operationally. This pattern motivates, but does not validate, early investment in executable foundations.

The selective flipped format and structured activities (Sections~\ref{subsec:lecture} and~\ref{sec:participation}) were manageable at this cohort size. The instructor used quizzes to surface misconceptions and activities to create immediate application opportunities. Feedback was consistent with this use, but the course did not compare it with an unflipped condition.

The course project supplied the primary synthesis mechanism (Section~\ref{sec:project}). Early proposal vetting addressed scope, while open-ended implementation exposed tooling and empirical challenges. Mixed-level teams created opportunities for peer interaction, and one project continued into a peer-reviewed paper; this follow-on case is illustrative rather than a general outcome measure.

Browser-based notebooks and open-source frameworks (Section~\ref{subsec:lecture}) avoided local environment setup. Exposure to multiple frameworks and execution backends also made ecosystem trade-offs available for comparison.
Additional tooling exploration in projects introduced friction from immature toolchains, evolving APIs, and ecosystem fragmentation, providing concrete QSE constraints for students to document.

\subsection{Challenges and Limitations}
\label{subsec:challenges}

The primary conceptual challenge was not mathematical preparation or probabilistic reasoning in isolation, but adapting to quantum algorithms as a fundamentally different algorithmic paradigm. While students were comfortable with linear algebraic representations and empirical variability, reasoning about algorithms whose behaviour is probabilistic, non-deterministic at the outcome level, and constrained by quantum information principles required explicit instructional scaffolding. Addressing this shift required flexibility in pacing and the use of execution-based examples that made algorithmic intent and correctness criteria concrete.

Managing cognitive load remained challenging, especially in lectures covering quantum algorithms and dense research literature. Trade-offs between rigour and accessibility were unavoidable given limited time. Some material was necessarily compressed, suggesting that future iterations could reallocate time based on student background or course positioning (further discussed in Section~\ref{sec:actionable}).

The use of multiple-choice questions in the undergraduate exam supported grading robustness and scalability but limited insight into students' reasoning. Correct answers did not always reveal whether understanding was partial or well grounded. Future iterations may therefore benefit from including a small number of carefully scoped free-form questions to better capture student reasoning while preserving grading feasibility.

The intentionally conservative assignment difficulty limited differentiation between levels of mastery. This reflected uncertainty about incoming proficiency in the first offering. Future offerings could retain the structure while increasing technical depth or strengthening integration with project work.

A recurring challenge in the course project was scope management, with many teams initially proposing overly ambitious goals. Iterative refinement during proposal review proved pedagogically valuable, enabling explicit discussion of feasibility, project sizing, and incremental development~---~skills directly relevant to SE practice.

The first offering was a moderate-size elective, not a mass-enrollment course. A larger offering could test moving activities to TA-led labs and adapting projects and presentations (Sections~\ref{sec:research_paper}~--~\ref{sec:participation}). The availability and preparation of TAs with sufficient quantum computing and QSE background would be a practical constraint; scalability was not evaluated here.

\subsection{Actionable Lessons for Instructors}
\label{sec:actionable}

The following considerations may guide adaptation, subject to the limitations in Section~\ref{sec:threats}.

The course structure is modular and can be adapted to different contexts. Foundational quantum material can be reduced for students with prior exposure, allowing deeper coverage of QSE topics. Similarly, modules such as PQC can be omitted or replaced without compromising overall coherence (see below).

The PQC module (Section~\ref{subsec:lecture}) was used as a self-contained boundary case rather than core QSE content. It illustrated how quantum computing affects classical systems through standards evolution, migration planning, backward compatibility, and risk management.
In practice, PQC was the only course component that substantially relied on prior computer security knowledge. This suggests that PQC can be treated as an optional or replaceable module, depending on instructional goals. If omitted, the computer security prerequisite can likewise be removed without compromising the coherence or learning objectives of the core QSE curriculum.

Treating Lecture~6 (PQC) as modular has direct implications for time allocation. Omitting this lecture would free a full three-hour block that could be reassigned to deeper coverage of QSE topics or to additional quantum algorithmic foundations, depending on instructional goals. Likewise, the lecture on the future of quantum computing (Lecture~10), while valuable for reflection, could be replaced with an application-focused module (e.g., optimization or quantum machine learning) to better align with student interests and support more hands-on or empirically grounded work. Finally, the recap and discussion lecture (Lecture~11) exhibited some flexibility in practice; with careful structuring, limited new material could be integrated into this session if needed. Taken together, these observations indicate that the lecture sequence contains latent flexibility that can be strategically reallocated to increase time devoted to core QSE or quantum operational concepts without expanding the overall course footprint.

In this offering, projects supplied sustained contact with tooling, empirical evaluation, and trade-offs that lectures alone did not provide. Structured proposals and iterative feedback were important for managing scope and feasibility.

A recurring challenge was fairly assessing individual work amid widespread LLM use. Individual assignments were therefore capped at 30\% of the final grade, with holistic grading for Assignment~2's open-ended writing. Similar courses should define expectations around tool use and authorship before assigning high weight to such work.

Embedding hands-on activities in the first lecture offering allowed direct observation of misconceptions and scaffolding needs. If enrolment increases, revised activities could be piloted in TA-led labs; students who completed the course may form part of the future TA pool.

While core quantum information concepts and SE principles are relatively stable, tooling ecosystems and QSE research evolve rapidly: according to~\cite[Fig. 2]{murillo2025quantum}, the number of QSE-related publications grew from fewer than 10 per year before 2019 to more than 200 in 2023. Course design should therefore anticipate change. 
Regular content review, reliance on openly available instructional materials, and use of latest research papers can facilitate timely updates while preserving conceptual continuity across course iterations.

Overall, this offering establishes that the documented course design was deliverable within one CS curriculum. Its adaptation and effects in other settings remain to be evaluated.

\section{Threats to Validity and Limitations}
\label{sec:threats}

This experience report has several limitations that should be considered when interpreting its observations and lessons.

First, the course was offered at a single institution and is based on a single delivery. As a result, the findings reflect a specific institutional context, student population, and instructor background. Outcomes may differ in other settings depending on curricular structure, prerequisite preparation, cohort composition, and local constraints.

Second, the evidence is primarily observational and qualitative. The course used no pre-test/post-test design, validated learning-outcomes instrument, control or comparison group, or inferential analysis of achievement. All instructional observations reflect the perspective of the course instructor, who was also the designer. Feedback, survey responses, and student work describe course dynamics but do not support causal claims.

Third, the course was intentionally designed for a mixed undergraduate--graduate audience. This heterogeneity complicates interpretation because academic level, background, and learning objectives may have influenced engagement, assessment, and classroom dynamics. Some observations may not apply to a single-level course.

Fourth, the elective nature of the course introduces selection effects. Students self-selected into the offering and were generally highly motivated and interested in the topic. Courses positioned as required components or aimed at broader student populations may encounter different challenges related to engagement, pacing, and workload.

Finally, because the course reflects a specific snapshot of a rapidly evolving field, some observations (particularly those related to concrete tools, platforms, or ecosystems) may have limited temporal validity. This limitation is inherent to experience reports in emerging technical domains and should be considered when interpreting the findings.

Taken together, these limitations suggest findings should be interpreted as context-sensitive insights rather than prescriptions. The goal of this experience report is to inform replication and adaptation, not claim universal effectiveness.

\section{Conclusion and Future Directions}
\label{sec:summary}

This paper reported the design and first offering of a QSE course that frames quantum computing as an SE problem domain. The course was delivered within one CS curriculum, and instructor observations and inspected artifacts were consistent with students engaging in executable, empirical, and software-system reasoning. The project supplied a setting for tooling constraints, probabilistic behaviour, reproducibility, and engineering trade-offs. These observations do not establish learning effects.

The experience also highlights directions for refinement and replication. Future iterations will strengthen early support for probabilistic reasoning, modestly increase assignment depth, and explore alternative formats for project presentations to improve scalability. More broadly, this experience suggests that QSE can be taught as a natural extension of SE, provided that courses explicitly address where classical assumptions break down and how they can be adapted. This report is intended to support further experimentation and adoption of QSE-focused offerings in SE education.

\bibliographystyle{IEEEtranN}
\bibliography{references}

\ifarxiv
\clearpage

\appendices

\section{Lecture-by-Lecture Preparation}
\label{sec:lecture_details}

This appendix summarizes the preparatory materials and conceptual focus associated with each lecture. The intent is to support reproducibility of the course design while keeping the main body of the paper focused on pedagogical rationale and lessons learned. Overviews below are descriptive rather than evaluative and reflect the structure of the first course offering. Table~\ref{tab:lecture_se_mapping} in the main paper maps the lecture sequence to the SE concerns emphasized in the course. For additional guidance, Figure~\ref{fig:course_flow} provides a high-level schematic of the course progression from quantum foundations to QSE topics.

\begin{figure}[htb]
\centering
\begin{tikzpicture}[
    node distance=0.4cm,
    box/.style={
        draw,
        rounded corners,
        align=center,
        font=\footnotesize,
        minimum width=0.8\columnwidth,
        minimum height=0.6cm,
        inner sep=2pt
    },
    arrow/.style={
        -{Latex[length=1.8mm]},
        thick
    }
]

\node[box] (entry) {Student entry: little or no prior quantum background};
\node[box, below=of entry] (fund) {Quantum information and circuits};
\node[box, below=of fund] (alg) {Quantum algorithms};
\node[box, below=of alg] (qse) {PQC and QSE topics};
\node[box, below=of qse] (synth) {Project-based synthesis and final assessment};

\draw[arrow] (entry) -- (fund);
\draw[arrow] (fund) -- (alg);
\draw[arrow] (alg) -- (qse);
\draw[arrow] (qse) -- (synth);

\end{tikzpicture}
\caption{High-level structure of the course. The sequence first builds foundational quantum-computing background, then transitions to post-quantum cryptography (PQC) and QSE topics, culminating in project-based synthesis and final assessment. Throughout the course, students engaged with executable artifacts, hands-on activities, empirical reasoning, and software-engineering concerns such as abstraction, testing, tooling, lifecycle, and migration.}
\label{fig:course_flow}
\end{figure}

\subsection{Lecture 1: Introduction and Motivation}
No preparatory materials were assigned. The lecture introduced the historical context of quantum computing, its anticipated applications, and common forecasts regarding performance, data-loading challenges, and applicability. It also outlined SE challenges arising both in legacy systems affected by quantum computing and in emerging quantum software.

\subsection{Lectures 2--3: Basics of Quantum Information}
Preparatory materials were assigned for these lectures. Topics included quantum bits, superposition, entanglement, unitary operators, and quantum circuits, with quantum teleportation used as a motivating example. The preparation emphasized conceptual familiarity with quantum state representation and evolution as a foundation for later algorithmic and engineering discussions.

Preparatory materials for \textit{Lecture~2} included an article by \smcitet{aaronson2021makes}, which discusses why quantum computing is difficult to explain and why it can appear counterintuitive to computer scientists. Students were also assigned Section~3 of \smcitet{fortnow2003one}, which presents a complexity-theoretic perspective on quantum computing and highlights differences in notation and reasoning styles. These readings were selected to provide contextual grounding and to normalize the perceived difficulty of the subject.

The core technical preparation consisted of the first two lessons from Unit~1 (``Basics of Quantum Information'') of the course by \smcitet{watrous2025understanding}, covering single and multiple quantum systems. These materials are mathematically substantive and were used to ensure that the course's introductory character did not come at the expense of conceptual rigour.

Preparatory materials for \textit{Lecture~3} continued with Unit~1 of \smcite{watrous2025understanding}. Lessons~3 and~4, focusing on quantum circuits and operational manifestations of entanglement, were assigned. These materials introduced circuit-based representations of quantum computation and completed the foundational unit on single and multiple quantum systems.

\subsection{Lectures 4--5: Quantum Algorithms}
Preparatory materials were provided selectively. These lectures introduced canonical quantum algorithms and the challenge of loading classical data into quantum computers. Algorithms such as Grover's~\smcite{grover1996fast} and Shor's~\smcite{shor1994algorithms} were discussed at different levels of abstraction, with emphasis on execution models and implementation considerations, and SE concerns such as abstraction boundaries, inspectability, and reasoning about algorithmic behaviour in executable form rather than through full formal derivations.

Preparatory work for \textit{Lecture~4} was limited\footnote{Although Unit~2 of \smcite{watrous2025understanding} provides a comprehensive treatment of quantum algorithms, its scope and depth make it impractical to cover fully within a six-hour lecture block.} to Lesson~6 of Unit~2 (``Quantum Algorithmic Foundations'') from \smcite{watrous2025understanding}. This lesson revisits classical computational models and Boolean circuits to establish a familiar baseline before introducing quantum algorithms, thereby helping students connect new quantum material to prior experience with algorithm structure and executable representations.

No preparatory materials were assigned for \textit{Lecture~5}. Given the cumulative cognitive load associated with quantum algorithms, this lecture relied on instructor-developed slides to present Grover's and Shor's algorithms at carefully chosen abstraction levels, prioritizing operational understanding, abstraction management, and engineering trade-offs. In particular, the lecture treated these algorithms not only as mathematical results but also as software artifacts that can be inspected, reasoned about, and discussed in terms of implementation structure, reuse, and black-box versus white-box understanding.

\subsection{Lecture 6: Post-Quantum Cryptography}
No formal preparatory materials were assigned. Instead, this lecture was assembled as a curated compilation of standards documents, research papers, and practitioner-oriented resources cited below. It introduced the impact of quantum computing on symmetric and asymmetric cryptography~\smcite{zhang2021quantum}, lattice-based cryptographic schemes~\smcite{bos2018crystals,ducas2018crystals}, and standardization efforts~\smcite{nist-fips-203,nist-fips-204,nist-fips-205}. It also addressed migration timelines~\smcite{nist-ir-8547} and SE challenges associated with cryptographic transitions~\smcite{zhang2023making}.

\subsection{Lectures 7--9: Introduction to QSE}
Preparatory materials were assigned for these lectures. Lecture~7 focused on the scope and foundations of QSE and its relationship to the software development lifecycle. Lecture~8 addressed quality assurance and service-oriented perspectives, while Lecture~9 focused on programming paradigms and language-level abstractions for quantum software.

Preparatory materials for \textit{Lecture~7} consisted of Sections~1 and~3 of the QSE roadmap by \smcitet{murillo2025quantum}, which provide an overview of motivations, scope, and research directions in the field.
This source was particularly useful instructionally because it is organized not merely as a literature review, but as a structured map of the main challenges in QSE together with representative work addressing them. As a result, it supports both broad orientation to the field and selective deeper exploration of individual subareas.

Preparatory materials for Lectures~8 and~9 were drawn from Section~4 of \smcitet{murillo2025quantum}. \textit{Lecture~8} used Sections~4.1 (Quality Assurance) and~4.2 (Quantum Service-Oriented Computing), while \textit{Lecture~9} focused on Section~4.4 (Programming Paradigms). In practice, this structure made lecture preparation relatively straightforward: an instructor could first use the roadmap to introduce a subdomain at a high level by summarizing its motivating challenges and the main lines of existing work, and then drill more deeply into selected topics, papers, or examples depending on instructional priorities and available time.

The modular structure of Section~4 in \smcitet{murillo2025quantum} allows alternative subdomains (e.g., model-driven engineering, software architecture, or software development processes) to be substituted depending on instructional priorities. In this sense, the roadmap functions not only as background reading, but also as a reusable scaffold for designing QSE lectures. For instructors preparing similar lectures, much of the high-level structure is therefore already available in the source itself: the main task is to select the subdomains most relevant to the course, present their motivating challenges, and decide where deeper coverage is pedagogically worthwhile.

\subsection{Lecture 10: The Future of Quantum Computing}
No preparatory materials were assigned. This lecture adopted a reflective and forward-looking perspective, using historical technology forecasts to discuss uncertainty, prediction, and long-term planning in quantum computing and related SE tasks.

\subsection{Lecture 11: Recap and Discussion}
This lecture was devoted to reviewing material from previous lectures and addressing student questions. No preparatory materials were required.

\subsection{Lecture 12: Project Presentations}
Students prepared short project presentations focusing on project motivation, methodology, and results. No additional preparatory readings were assigned.

\section{In-Class Participation Activities}
\label{sec:participation_activities}

This appendix provides representative examples of the in-class participation activities conducted during Lectures~2 through~8. The activities are described to support reproducibility of the course design. Assessment mechanics and pedagogical rationale are discussed in Section~\ref{sec:participation}.

\subsection{Lecture 2: Quantum State Transformation and Noise Simulation}
Students implemented basic quantum state transformations using NumPy~\smcite{harris2020array}, representing quantum states as vectors and quantum gates as matrices. By applying matrix-vector multiplication explicitly, students observed state evolution under gate application. A simple noise model used pseudo-random perturbations to illustrate non-ideal behaviour and enable discussion of how noise can be modeled and analyzed.

\subsection{Lecture 3: Quantum Teleportation and Language Expressivity}
Students implemented the quantum teleportation protocol using a reference protocol inspired by~\smcitet{watrous2025understanding} within a circuit-based framework. The activity included exporting the circuit to quantum assembly and importing it into another programming framework.
This workflow exposed architectural and language-level constraints, particularly limitations due to the absence of real-time classical control flow in some toolchains at the time the course was conducted.
Students then implemented teleportation natively in a second framework, enabling comparison of language expressivity, abstraction choices, and representational constraints.

\subsection{Lecture 4: Backend Comparison and Black-Box Algorithms}
Students measured Bell states across multiple execution environments, including ideal simulators, noisy simulators, and real quantum hardware. Outcomes were compared to illustrate variability across backends. A second exercise examined a black-box implementation of Shor's algorithm from a deprecated quantum algorithm library~\smcite{qiskit_aqua}. Students inspected the source code, thereby treating the implementation as a white box, and discussed its structure, evolution, and deprecation from an SE perspective.

\subsection{Lecture 5: White-Box and Black-Box Grover Implementations}
Students completed a partially implemented version of Grover's algorithm, experimented with iteration counts, and observed performance trends. This white-box implementation was contrasted with a packaged black-box version of the same algorithm. The comparison prompted discussion of abstraction boundaries, reuse, and testing considerations. The session concluded by mapping Grover's algorithm to constraint-satisfaction problems such as 3-SAT~\smcite{cook1971complexity}.

\subsection{Lecture 6: PQC Role-Play}
Participation took the form of a role-playing exercise in which groups assumed the roles of security architects, attackers, and compliance officers. Teams discussed PQC migration scenarios, identified risks and threat models, and proposed mitigation strategies while considering regulatory and compliance constraints. The exercise foregrounded cross-cutting SE concerns rather than cryptographic detail.

\subsection{Lecture 7: Quantum Software Development Lifecycle Challenge}
The class was divided into teams aligned with different phases of the software development lifecycle, including requirements engineering, design, service-oriented computing, and process management. Each team produced artifacts relevant to its assigned phase for a hypothetical quantum-enhanced system. Outputs were then discussed to examine consistency, dependencies, and gaps across lifecycle stages.

\subsection{Lecture 8: Debugging and Testing Quantum Programs}
Students performed a structured debugging exercise on a provided quantum program snippet. Groups identified defects, proposed testing strategies, and discussed possible fixes. The subsequent discussion compared alternative debugging approaches and highlighted challenges specific to testing quantum software.

\section{Representative Assignment Materials}
\label{sec:assessment_materials}
This appendix details the course assignments.

\subsection{Assignment 1: Skills Bootstrapping}
The first assignment combined an externally administered fundamentals test with small implementation tasks on classical data encoding for quantum computation. Students were required to:
\begin{enumerate}
    \item pass the ``Basics of Quantum Information'' test offered as part of~\smcite{watrous2025understanding};
    \item implement basis encoding for example bitstrings and submit code together with both state-preparation and transpiled circuits;
    \item answer conceptual questions about duplicate bitstrings and sparse representations for large binary datasets; and
    \item implement amplitude encoding for a small numeric dataset and answer conceptual questions about handling negative values.
\end{enumerate}

Representative prompts included:
\begin{itemize}
    \item Suppose your dataset contains the bitstrings \texttt{0101} and \texttt{1011}. Encode them using basis encoding. Submit the corresponding code and images of both the state-preparation and transpiled circuits.
    \item Suppose we need to encode ten bitstrings, each $128$ bits long. This would require a statevector of length $2^{128}$, which is infeasible to store directly. How would you approach this task on a computer with only 16GB of RAM? Consider sparsity.
    \item Suppose your dataset contains the values $[2, 7, 6, 2, 5]$. Encode them using amplitude encoding. Download the resulting circuits, including both the state preparation and transpiled circuit. Use \textsc{rx}, \textsc{ry}, \textsc{rz}, and \textsc{cx} gates as basis gates. Submit the corresponding code and images. How would you approach encoding a dataset that includes negative values? How would you preserve information about negative numbers?
\end{itemize}

Assignment~1 was assessed primarily for completion and technical correctness of the required encoding tasks, together with the quality of responses to the associated conceptual questions.

\subsection{Assignment 2: Conceptual Synthesis in QSE}
The second assignment required short analytical responses (approximately 0.5--1 page per question) intended to synthesize course material across multiple lectures. Students submitted a PDF responding to four prompts, each weighted equally:
\begin{itemize}
    \item analysis of the NIST post-quantum cryptography standardization timeline and its implications for industry, including challenges of migration and a timeline-based upgrade strategy;
    \item comparison of early and later definitions of QSE (shown in~\smcite[Sec. 3.2]{zhao2020quantum}), together with discussion of what those definitions imply about the evolution of quantum hardware and hybrid quantum--classical software development;
    \item comparison of quantum simulators and real quantum hardware for software testing, including advantages, limitations, and the impact of noise, resource constraints, and scalability;
    \item comparison of classical and quantum programming paradigms, including discussion of quantum-native data types and higher-level abstractions that could better support software design.
\end{itemize}

 Assignment~2 was assessed holistically, emphasizing clarity of argumentation, technical correctness, conceptual synthesis, and the ability to connect software engineering concerns with quantum-specific constraints.

\section{Project Assessment and Deliverables}
\label{sec:project_appendix}

This appendix summarizes the assessment artifacts and expectations associated with the course project. The intent is to make the project structure easier to replicate or adapt in future offerings.

\subsection{Project Proposal}
Projects began with an early proposal submitted in the form of a structured abstract. Required sections were ``Background'', ``Objective'', ``Methods'', ``Expected Results'', ``Conclusion'', and ``Implications''. For multi-person teams, an additional ``Roles'' section identified the expected contributions and responsibilities of each team member. To encourage early scope control and clear planning, each section was expected to remain concise.

\subsection{Final Project Report and Software Artifacts}
The final deliverable was a written report submitted in PDF format using the ACM article template. The report was limited to 10 pages, excluding references and appendices. Expected sections included a title page, abstract, introduction, methodology, results and discussion, conclusion and future work, and references. A literature review and appendices were included when appropriate. Reports were expected to describe methods in sufficient detail to support reproducibility and to critically analyze results rather than merely report outcomes.

Projects involving software artifacts were also expected to include source code together with a short \texttt{README.md} file describing how to run the implementation. Supplementary materials, such as datasets, notebook files, or circuit descriptions, could be submitted separately when needed.

\subsection{Project Presentation}
Each team delivered a short in-class presentation during the final lecture. Presentations were limited to five minutes and were expected to address three questions: what the project does, why it is important, and what high-level results were obtained. Slides were normally submitted in PDF format, unless another format was required for technical reasons such as animation or live demonstration.

\subsection{Evaluation Criteria}
The project proposal was evaluated for clarity of background and objective, appropriateness of methods, plausibility of expected results, strength of implications, and overall structure. The final report was evaluated for technical depth, organization and clarity, conclusion and future-work discussion, adherence to formatting expectations, use of references and supporting material, creativity and originality, and presentation of visuals or supplementary artifacts. Together, these criteria balanced open-ended exploration with disciplined execution.

\section{Representative Formative Feedback Instruments}
\label{sec:feedback_forms}

This appendix summarizes representative feedback instruments used during the course to collect formative input on lecture pacing, clarity, workload, and hands-on activities. The intent of these forms was not formal educational evaluation, but rapid instructional calibration during the first offering.

\subsection{Weekly Lecture Feedback}
A short anonymous weekly form was used to collect feedback on the lecture experience. Representative items included:
\begin{itemize}
    \item course pace (too fast / just right / too slow),
    \item amount of material presented (too much / just right / not enough),
    \item amount of interaction during the lecture (too much / just right / not enough),
    \item whether the lecture was useful,
    \item whether the instructor explained the material clearly, and
    \item an open-ended suggestions/comments field.
\end{itemize}

The purpose of this form was to identify whether pacing, density, or interaction style required adjustment from week to week.

\subsection{Hands-on Activity Feedback}
A separate anonymous form was used to collect feedback on the hands-on component of selected lectures. Representative items included:
\begin{itemize}
    \item overall experience with the hands-on activity,
    \item ease of access to the lab materials (e.g., via Google Colab),
    \item whether technical difficulties were encountered,
    \item ease of submission through the learning-management system,
    \item whether the activity improved understanding of the course material,
    \item clarity of the lab instructions,
    \item sufficiency of the time allocated for the activity,
    \item open-ended questions about confusing or challenging aspects of the lab, and
    \item open-ended suggestions for improvement.
\end{itemize}

These responses were used to refine the balance between explanation time, independent exploration time, and submission logistics, and to identify where additional scaffolding or clearer instructions were needed.

\section{Representative Final Exam Questions}
\label{sec:exam_examples}

This appendix provides representative examples of undergraduate final-exam questions. The intent is to illustrate the breadth of concepts assessed, spanning foundational quantum computing, PQC, and QSE.

\subsection{Quantum Foundations}
\begin{itemize}
    \item What is the dimension of the Hilbert space (statevector) for a 3-qubit system?
    \begin{enumerate}[label=\Alph*., leftmargin=2em]
        \item 3
        \item 6
        \item 8
        \item 16
    \end{enumerate}
    Correct answer: \textbf{C. 8}

    \item Which of the following states is not a product state (i.e., is entangled)?
    \begin{enumerate}[label=\Alph*., leftmargin=2em]
        \item $\frac{1}{\sqrt{2}}\left(\ket{00} + \ket{11}\right)$
        \item $\ket{0} \otimes \ket{1}$
        \item $\left(\frac{1}{\sqrt{2}}(\ket{0} + \ket{1})\right)\otimes\ket{0}$
        \item $\ket{1}\otimes\left(\frac{1}{\sqrt{2}}(\ket{0} + \ket{1})\right)$
    \end{enumerate}
    Correct answer: \textbf{A. $\frac{1}{\sqrt{2}}\left(\ket{00} + \ket{11}\right)$}
\end{itemize}

\subsection{Security and Post-Quantum Cryptography}
\begin{itemize}
    \item Grover's algorithm affects symmetric encryption by
    \begin{enumerate}[label=\Alph*., leftmargin=2em]
        \item giving only a quadratic speed-up, so doubling key length restores security
        \item giving an exponential speed-up, so any key length is broken
        \item factoring private keys
        \item providing perfect secrecy
    \end{enumerate}
    Correct answer: \textbf{A. giving only a quadratic speed-up, so doubling key length restores security}

    \item The ``Harvest-and-Decrypt'' attack refers to
    \begin{enumerate}[label=\Alph*., leftmargin=2em]
        \item stealing private keys from RAM
        \item collecting ciphertext now to decrypt later with quantum computers
        \item inserting malicious quantum hardware
        \item exploiting speculative execution
    \end{enumerate}
    Correct answer: \textbf{B. collecting ciphertext now to decrypt later with quantum computers}
\end{itemize}

\subsection{Quantum Software Engineering}
\begin{itemize}
    \item Statistical assertions rely on \_\_\_\_ to reconstruct quantum states.
    \begin{enumerate}[label=\Alph*., leftmargin=2em]
        \item gate synthesis
        \item hypothesis testing
        \item error-correcting codes
        \item classical compilation
    \end{enumerate}
    Correct answer: \textbf{B. hypothesis testing}

    \item Quantum noise is cited as a chief obstacle to straightforward error detection in quantum programs.
    \begin{enumerate}[label=\Alph*., leftmargin=2em]
        \item True
        \item False
    \end{enumerate}
    Correct answer: \textbf{A. True}

    \item Current quantum languages generally operate at a low abstraction level, exposing gate primitives directly.
    \begin{enumerate}[label=\Alph*., leftmargin=2em]
        \item True
        \item False
    \end{enumerate}
    Correct answer: \textbf{A. True}
\end{itemize}
\fi

\end{document}